\documentclass{article}[12pt]

\def \lf  {\left (}
\def \rt  {\right )}

\begin{document}

\thispagestyle{empty}

\begin{center}

{\hbox to \hsize {\hfill CU-TP-1114}}

\bigskip

{\Large \bf A Secret Tunnel Through The Horizon}

\bigskip
\bigskip
\bigskip

{\large \sc Maulik~Parikh\footnote{\tt mkp@phys.columbia.edu}}

{\it Department of Physics, Columbia University, New York, NY
  10027}

\vspace*{1.5cm}

\large{\bf Abstract}
\end{center}
\noindent
Hawking radiation is often intuitively visualized as particles that have
tunneled across the horizon. Yet, at first sight, it is not
apparent where the barrier is. Here I show that the barrier depends on
the tunneling particle itself. The key is to implement energy
conservation, so that the black hole contracts during the process of
radiation. A direct consequence is that the radiation spectrum cannot
be strictly thermal. The correction to the thermal spectrum is of
precisely the form that one would expect from an underlying unitary quantum
theory. This may have profound implications for the black hole
information puzzle.
\bigskip
\bigskip

\begin{center}
\noindent
{\em This essay was awarded First Prize in the 2004 Essay Competition
  of the Gravity Research Foundation.}
\end{center}

\newpage

\setcounter{page}{1}

\noindent
Classically, a black hole is the ultimate prison: anything that enters
is doomed; there is no escape. Moreover, since nothing can
ever come out, a classical black hole can only grow bigger with
time. Thus it came as a huge shock to physicists when Stephen Hawking
demonstrated that, quantum mechanically, black holes could actually radiate
particles. With the emission of Hawking radiation, black holes
could lose energy, shrink, and eventually evaporate completely.

How does this happen? When an object that is classically stable
becomes quantum-mechanically unstable, it is natural to suspect
tunneling. Indeed, when Hawking first proved the existence of
black hole radiation \cite{hawking}, he described it as 
tunneling triggered by vacuum fluctations near the
horizon. The idea is that when a virtual particle pair is created just inside
the horizon, the positive energy virtual particle can tunnel out -- no
classical escape route exists -- where it materializes as a real
particle. Alternatively, for a pair created just outside the horizon,
the negative energy virtual particle, which is forbidden outside, can
tunnel inwards. In either case, the negative energy
particle is absorbed by the black hole, resulting in a decrease in the
mass of the black hole, while the positive energy particle escapes to
infinity, appearing as Hawking radiation.

This heuristic picture has obvious visual and intuitive appeal. But, oddly,
actual derivations of Hawking radiation did not proceed in this way at
all \cite{hawking,gibbonshawk}. There were two apparent hurdles. The
first was technical: in order to do a tunneling computation one needed
to have a coordinate system that was well-behaved at the horizon; none
of the well-known coordinate systems were suitable. The second hurdle
was conceptual: there didn't seem to be any barrier!  Typically,
whenever a tunneling event takes place, there are two separated
classical turning points which are joined by a trajectory in imaginary
or complex time. In the WKB or geometrical optics limit, the
probability of tunneling is related to the imaginary part of the
action for the classically forbidden trajectory via
\begin{equation}
\Gamma \sim \exp (-2 ~{\rm Im}~ I) \; ,
\end{equation}
where $I$ is the action for the trajectory. So, for example, in
Schwinger pair production in an electric field, the action for the
trajectory that takes the electron-positron pair to their required
separation yields the rate of production. Now, the problem with black
hole radiation is that if a particle is even infinitesimally outside
the horizon, it can escape classically. The turning points therefore seem
to have zero separation, and so it's not immediately clear what
joining trajectory is to be considered. What sets the scale for tunneling?
Where is the barrier? 

In this essay, I will show that the intuitive picture is more than a picture:
particles do tunnel out of a black hole, much as Hawking had first
imagined. But they do this in a rather subtle way since, as just
argued, there is no pre-existing barrier. Instead, what happens is
that the barrier is created by the outgoing particle itself. The
crucial point is that energy must be conserved \cite{energy}. As the
black hole radiates, it loses energy. For black holes, the energy and
radius are related, and this means that the black hole has to shrink. It is
this contraction that sets the scale: the horizon recedes from its
original radius to a new, smaller radius. Moreover, the amount of
contraction depends on the energy of the outgoing particle so, in a
sense, it is the tunneling particle itself that secretly defines the
barrier.

Now, one might fear that a calculation of Hawking radiation in which
energy conservation is critical would require a quantum theory of
gravity because the metric must fluctuate to account for the contraction
of the hole. This is true but, fortunately, there is
at least one regime in which gravitational back-reaction can be accounted for
reliably and that is the truncation to spherical
symmetry. Intuitively, in a transition from one
spherically symmetric configuration to another, no graviton is emitted
because the graviton has spin two and each of the spherically
symmetric configurations has spin zero. So quantizing a spherically symmetric
matter-gravity system is possible because no quantization of
gravitons is required. Indeed, the only degree of freedom is the
position of the particle (which, being spherically symmetric, is
actually a shell).

Armed with these insights, we can compute the imaginary part of the
action for a particle to go from inside the black hole to outside. A
convenient line element for this purpose is
\begin{equation}
ds^2 = - \lf 1 - {2 M \over r} \rt dt^2 + 2 \sqrt{2 M \over r}
dt~ dr + dr^2 + r^2 d \Omega_2^2 \; .
\end{equation}
This line element was first written down by Painlev\'e and
Gullstrand long ago \cite{painleve,gullstrand}, but they apparently
missed the significance of their discovery. What they had unwittingly
found was a coordinate system that was well-behaved at the horizon.
Other nice features of their coordinate system are that
there is no explicit time dependence, and constant-time slices are
just flat Euclidean space.

Working in these coordinates within the spherically symmetric truncation,
one calculates the imaginary part of the action,
\begin{equation}
{\rm Im}~ I = {\rm Im} \int_{r_{\rm in}}^{r_{\rm out}} p~ dr \; ,
\end{equation}
where $p$ is the momentum, $r_{\rm in} = 2M$ is the initial radius of
the black hole, and $r_{\rm out} = 2(M - E)$ is the final radius of
the hole. Here $E$ is 
the energy of the outgoing particle. Notice that this fixes the scale:
the classical turning points, $2M$ and $2(M-E)$ are separated by an
amount that depends on the energy of the particle. It is the forbidden
region from $r = 2M$ to $r = 2(M-E)$ that the tunneling particle must
traverse. That's the barrier.

One would then expect that, in the WKB limit, the probability of
tunneling would take the form
\begin{equation}
\Gamma \sim \exp (- 2 ~{\rm Im}~ I) \approx \exp (-\beta E) \; ,
\label{gamma}
\end{equation}
where $e^{-\beta E}$ is the Boltzmann factor appropriate for an
object with inverse temperature $\beta$. Indeed, this is {\em almost}
what is found. But, remarkably, an exact calculation \cite{energy,tunnel} of
the action for a tunneling spherically symmetric particle yields
\begin{equation}
\Gamma \sim \exp \lf - 8 \pi M E \lf 1 - {E
  \over 2M } \rt \rt \; .  \label{rate}
\end{equation}
If one neglects the $E/2M$ term in the expression, it does take the
form $e^{-\beta E}$ with precisely the inverse of the temperature
that Hawking found. So at this level we have confirmed that Hawking
radiation can be viewed as tunneling particles and, furthermore, we
have verified Hawking's thermal formula. But, unlike traditional
derivations, we have also taken into account the conservation of
energy and this yields a correction, the additional term $E/2M$. Thus
the spectrum is not precisely thermal!

This is exciting news because arguments that information is lost
during black hole evaporation rely in part on the assumption of strict
thermality of the spectrum \cite{infoloss}. That the spectrum is not
precisely thermal may open the way to looking for information-carrying
correlations in the spectrum -- work on this continues. Indeed, the
exact expression including the $E/2M$ term can be cast rather
intriguingly in the form 
\begin{equation}
\Gamma \sim \exp(\Delta S) \; , \label{deltaS}
\end{equation}
where $\Delta S$ is the change in the Bekenstein-Hawking entropy of
the hole. This is a very interesting form for the answer to take for a
number of reasons, including the fact that it is consistent with
unitarity. Put another way, our result agrees exactly with what we would
expect from a quantum-mechanical microscopic theory of black
holes in which there is no information loss! For quantum theory
teaches us that the rate for a process is expressible as the square of
the amplitude multiplied by the phase space factor. In turn, the phase
space factor is obtained by summing over final states and averaging
over initial states. But, for a black hole, the number of such states
is just given by the exponent of the final and initial Bekenstein-Hawking
entropy:
\begin{equation}
\Gamma = |{\rm amplitude}|^2 \times (\mbox{phase space factor}) \sim
       {e^{S_{\rm final}} \over e^{S_{\rm initial}}} = \exp(\Delta S) \; . 
\end{equation}
Quantum mechanics, we observe, is in perfect agreement with our answer.


\begin{thebibliography}{99}

\bibitem{hawking}
 S. W. Hawking, ``Particle Creation by Black Holes,''
 Commun. Math. Phys. {\bf 43} (1975) 199.

\bibitem{gibbonshawk}
G. W. Gibbons and S. W. Hawking, ``Action integrals and partition
functions in quantum gravity,'' Phys. Rev. {\bf D15} (1977) 2752.

\bibitem{energy}
M. K. Parikh, ``Energy Conservation and Hawking Radiation,'' {\tt
  hep-th/0402166}. 

\bibitem{painleve}
P. Painlev\'e, ``La m\'ecanique classique et la th\'eorie de la
relativit\'e,'' Compt. Rend. Acad. Sci. (Paris) {\bf 173} (1921) 677.

\bibitem{gullstrand}
A. Gullstrand, ``Allgemeine L\"osung des statischen
Eink\"orper-problems in der Einsteinschen Gravitationstheorie,''
Arkiv. Mat. Astron. Fys. {\bf 16} (1922) 1.

\bibitem{tunnel}
M. K. Parikh and F. Wilczek, ``Hawking Radiation as Tunneling,'' 
Phys. Rev. Lett. {\bf 85} (2000) 5042; {\tt hep-th/9907001}.

\bibitem{infoloss}
S. W. Hawking, ``Breakdown of Predictability in Gravitational
Collapse,'' Phys. Rev. {\bf D14} (1976) 2460.

\end{thebibliography}
\end{document}